## CONDENSED-MATTER SPECTROSCOPY

# Luminescence of a ZnO:Ga Crystal upon Excitation in Vacuum UV Region

**P. A. Rodnyi**[a], **G. B. Stryganyuk**[b], **and I. V. Khodyuk**[a]

[a] *St. Petersburg State Technical University, St. Petersburg, 195251 Russia*
[b] *Franko Lviv National University, Lviv, 79005 Ukraine*
Received March 29, 2007

**Abstract**—The spectral–kinetic characteristics of a ZnO:Ga single crystal upon excitation in the vacuum UV region have been studied. At a temperature of 8 K, the exciton luminescence line peaking at 3.356 eV has an extremely small half-width (7.2 meV) and a short decay time (360 ps). In the visible range, a wide luminescence band peaking at ~2.1 eV with a long luminescence time at 8 K and a decay time in the nanosecond range at 300 K is observed. The luminescence excitation spectra of ZnO:Ga have been measured in the range from 4 to 12.5 eV.

**DOI:** 10.1134/S0030400X08020100

Zinc oxide has been actively investigated since it has a number of interesting physical properties as a wide-gap ($E_g$ = 3.4376 eV at 4.2 K) semiconductor with a significant ionic-bonding component [1]. The prospects of application of ZnO and ZnO:Ga in short-wavelength optoelectronics and laser and scintillation technique were considered in [2–4]. Photoluminescence investigations, performed on thin films [2, 3], ceramics [4], nanopowders [5], and single crystals [6–8] showed that zinc oxide exhibits a narrow UV line peaking near 3.3 eV and a wide long-wavelength (~2.2 eV) emission band. The narrow UV line has an exciton nature, while the long-wavelength band is attributed to recombination of electrons with oxygen vacancies [5]. Most investigations of the ZnO photoluminescence were performed upon excitation by a He–Cd laser (325 nm) [5–8]. In this study, the spectral–kinetic characteristics of a ZnO:Ga single crystal excited by vacuum UV (VUV) radiation were investigated.

Measurements were performed on a small (approximately $4 \times 1.5 \times 1$ mm) ZnO:Ga single crystal with a gallium concentration of $10^{18}$ cm$^{-3}$. The excitation and emission spectra of ZnO:Ga were measured at temperatures of 8 and 300 K at the Synchrotron Radiation Laboratory HASYLAB (DESY, Hamburg), using the experimental equipment of the SUPERLUMI station. The luminescence spectra were recorded using a monochromator with the best resolution (for low-temperature measurements): 2 meV. The luminescence excitation spectra were measured in the range from 4 to 12.5 eV (resolution 3.2 Å) using an R6358P photomultiplier (Hamamatsu) on a secondary monochromator. In measurements of kinetic curves, the excitation pulse width was 290 ps.

Figure 1 shows the crystal luminescence spectra measured upon excitation by 6.89-eV photons at temperatures of 8 and 300 K (Figs. 1a and 1b, respectively). The low-temperature spectrum contains a very narrow (width at half-maximum 7.2 meV) line peaking at 3.356 eV (Fig. 1‡, inset) and a wide band peaking near 2.0 eV. The narrow 3.356-eV UV line is due to free excitons in zinc oxide. The weak peak at 3.214 eV should be attributed to the superposition of a free exciton and two longitudinal optical (LO) phonons, since the energy of LO phonons in ZnO is 72 meV [7]. The wide band is red-shifted in comparison with the similar band in ZnO nanocrystals. It is believed that this band is due to several types of electron recombination: at oxygen and zinc vacancies and at interstitial zinc [8].

At room temperature, the UV line is red-shifted ($h\nu_{max}$ = 3.3 eV) and broadened (Fig. 1b). These changes are due to the change in the luminescence mechanism: the exciton luminescence is suppressed at room temperature and emission occurs between donor (gallium) levels and valence-band holes [3, 9]. The wide band in the visible spectral range (Fig. 1b) peaks at 2.1 eV and has a feature near 2.3 eV.

The reflection and luminescence excitation spectra of a ZnO:Ga crystal are shown in Fig. 2. The excitation spectra exhibit peaks near 6.0, 7.8, and 11.4 eV. Generally, peaks in this region are due to the formation of high-energy excitons. UV and visible (Fig. 2, curves *1* and *2*, respectively) luminescence excitation bands have a similar shape; this fact is indicative of similar mechanisms of energy transfer to the corresponding luminescence centers. In the crystal reflection spectrum (Fig. 2, *3*), the high-energy peaks at 11.2 and 12.0 eV should be attributed to the core zinc states $3d_{5/2}$ and $3d_{3/2}$, respectively.

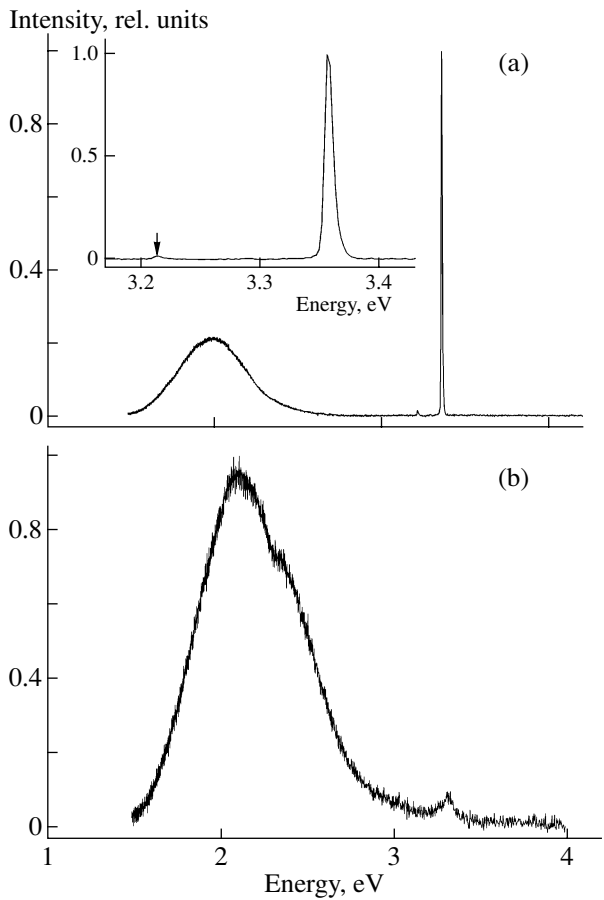

**Fig. 1.** Luminescence spectra of a ZnO:Ga crystal excited by 6.89-eV photons at (a) 8 and (b) 300 K. The inset shows the luminescence spectrum at 8 K on the enlarged energy scale.

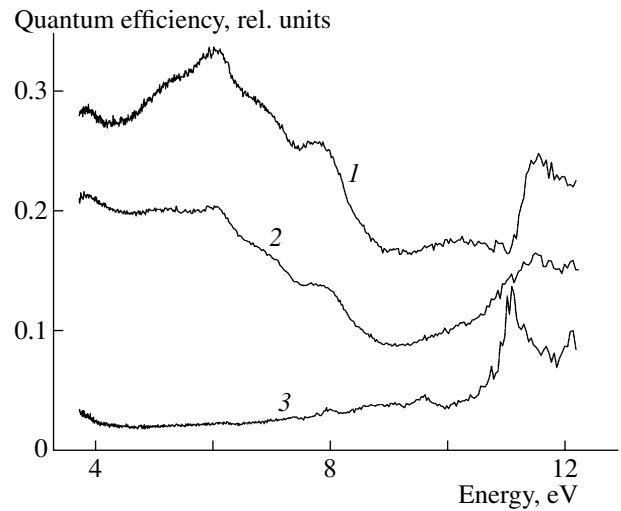

**Fig. 2.** Excitation spectra of (*1*) UV and (*2*) visible luminescence of a ZnO:Ga crystal at 8 K and (*3*) its reflection spectrum at 300 K.

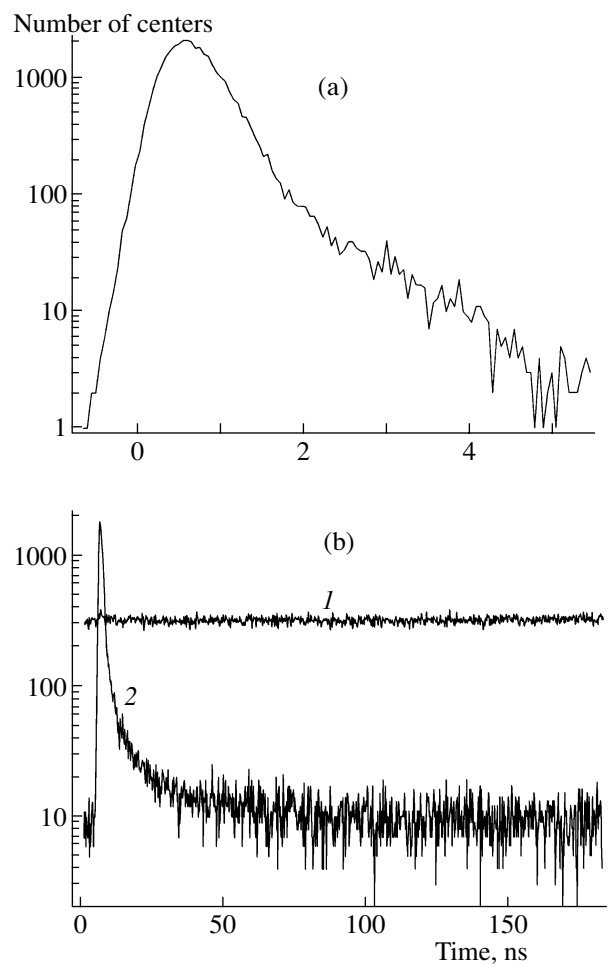

**Fig. 3.** Luminescence kinetics of a ZnO:Ga crystal: (a) the UV line at 8 K and (b) the visible band at temperatures of (*1*) 8 and (*2*) 300 K.

The kinetics of UV and visible luminescence upon VUV excitation is shown in Fig. 3. At 8 K, the main UV luminescence decay constant is 360 ps (Fig. 3a); this value is characteristic of excitons in ZnO. The decay time of the band at 2.0 eV is difficult to determine in our case, since it is in the microsecond range (Fig. 3b, *1*). At room temperature, the visible luminescence kinetics has a complex character; in this case, the luminescence intensity decreases by two orders of magnitude over ~20 ns (Fig. 3b, *2*). The significant decrease in the decay time of the visible luminescence at 300 K in comparison with that at 8 K indicates the presence of shallow traps in the crystal. At low temperatures, the release rate of carriers from traps is low, as a result of which the luminescence decay constant is large. With an increase in temperature, the carrier release rate increases, and, accordingly, the deexcitation time decreases. The possibility of forming shallow electron traps due to the introduction of Ga into ZnO was noted in [8].

Thus, the VUV-excitation luminescence spectra of ZnO:Ga crystals have a number of features, among which the most important ones are the extremely small width of the exciton luminescence line, a short decay time of this luminescence, and a large difference in the decay times of the visible luminescence band at low and high temperatures.